\documentclass{tWRM2e}
\usepackage[latin1]{inputenc}
\usepackage{amsmath}
\usepackage{amsfonts}
\usepackage{amssymb}
\usepackage{graphicx}
\usepackage{epsfig}
\usepackage{here}
\usepackage{subfigure}
\begin{document}
\title{Giant fluctuations of topological charge\\ in a disordered wave guide}
\date{\today}

\author{D. Anache-M\'enier$^{\ast}$\thanks{$^\ast$Corresponding author. Email: Domitille.Anache@grenoble.cnrs.fr
\vspace{6pt}} and B.A. van Tiggelen\\\vspace{6pt}  Laboratoire de Physique et Mod\'{e}lisation des
Milieux Condens\'{e}s, CNRS /Universit\'{e} Joseph Fourier, BP
166, F-38042 Grenoble Cedex 9, France}

\maketitle

% \author{D. Anache-M\'{e}nier and B. A. van Tiggelen}
% \affiliation{Laboratoire de Physique et Mod\'{e}lisation des Milieux Condens\'{e}s, CNRS /Universit\'{e} Joseph Fourier, BP 166, F-38042 Grenoble Cedex 9, France}
% %\author{B. A. van Tiggelen}
% \affiliation{Laboratoire de Physique et Mod\'{e}lisation des Milieux Condens\'{e}s, CNRS /Universit\'{e} Joseph Fourier, BP 166, F-38042 Grenoble Cedex 9, France}

 \begin{abstract}
 We study the fluctuations of the total topological charge of a scalar wave propagating in a hollow conducting wave guide filled with scatterers inside. We investigate the dependence of the screening on the scattering mean free path and on the presence of boundaries. Near the cut-off frequencies of the wave guide, screening is strongly suppressed near the boundaries. The resulting huge fluctuations of the total topological charge are very sensitive to the disorder.
\end{abstract}
% \abstract{We study the fluctuations of the total topological charge of a scalar wave propagating in a hollow conducting wave guide filled with scatterers inside. We investigate the dependence of the screening on the scattering mean free path and on the presence of boundaries. Near the cut-off frequencies of the wave guide, screening is strongly suppressed near the boundaries. The resulting huge fluctuations of the total topological charge are very sensitive to the disorder.}

%\begin{keywords}
%Phase singularities, Gaussian speckle, Wave guide, Fluctuations of charge.
%\end{keywords}

\section{Introduction}

A complex scalar random wave field is given by: $\Psi(\vec{r},t)=A(\vec{r},t)\exp(\mathrm i \Phi(\vec{r},t))=\chi(\vec{r},t)+\mathrm i\eta(\vec{r},t)$. When both its real part $\chi$ and its imaginary part $\eta$ cancel, the amplitude $A$ also cancels but the phase $\phi$ is left undefined: it is a phase singularity. In space, these singularities constitute nodal lines located at the intersection of the two surfaces defined by $\chi(\vec{r},t)=0$ and $\eta(\vec{r},t)=0$. On a flat surface inside the wave field, the phase singularities are points. The phase singularities show up at the intersection of equiphases. While turning around a singularity the phase changes by $2\pi p$ with $p\in \mathbb{Z}$; $q=\pm p$ is called the topological charge associated with the phase singularity and its sign is determined by the sign of the phase vortex. It is known that for Gaussian statistics of the field, large topological charges have a small probability so that we can restrict to $q=\pm1$ \cite{Berry78,Saddles}.\\

The total topological charge present on a surface $\mathcal{S}$ is defined as the sum of the charges of the singularities located on $\mathcal{S}$: $Q\equiv\sum_{n\in\mathcal{S}} q_n$. According to Stokes' theorem, the total topological charge $Q$ is also equal to the accumulated phase along the contour $\mathcal{C}$ of surface $\mathcal{S}$ : 
\begin{equation}\label{Stokes}
Q=\frac{1}{2\pi}\oint_\mathcal{C} d\mathbf{s}\cdot\boldsymbol{\nabla}\Phi.
\end{equation}
\noindent In this paper we study the statistics of $Q$ and their dependence on the degree of disorder in a wave guide.\\

In a random Gaussian speckle pattern generated by a $3D$ infinite medium, the density of singularities on a flat surface is $d=2\pi/3\lambda^2$ \cite{berry2}. Since near field speckle spots are typically $\lambda^2$ in size, each speckle spot  contains approximatively two singularities. As we average over the disorder, $<Q(\mathcal{S})>=0$  but the fluctuations of the total charge, quantified by the variance $<Q^2(\mathcal{S})>$, may be sensitive to both the surface $\mathcal{S}$ and the mean free path $\ell$ of the waves.\\

Let's consider the fluctuations $<Q^2(R)>$ of the total topological charge contained in a circular surface $\mathcal{S}=\pi R^2$ of radius $R$. One basic feature of $<Q^2(R)>$ is already known. If we would assume all  nodal points $n$ to have random charges $q_n=\pm1$, with equal probability and independent to each other, we would  find that $<Q^2(R)>=<(\sum_{n=1}^{N}q_n)^2>= N=\pi R^2 d$, i.e. the fluctuations are  proportional to the surface. However, this scenario turns out to be invalid, at least for infinite media. Indeed it has been shown that zeros with positive charge tend to be surrounded by zeros with negative charge and vice versa \cite{Shvartsman,twin,Screening}. Topological charges are not independent but tend to be screened, making the fluctuations grow slower than a $R^2$ quadradic law. This screening is similar to the one of electrical charge in ionic fluids and plasmas.\\

For a random Gaussian superposition of plane waves in space, Wilkinson and Freund \cite{freund} report a linear, ``diffuse'' asymptotic form: $<Q^2(R)>\rightarrow R$ for large $R$. Berry and Dennis \cite{berry2} use Gaussian-smoothed boundaries and show that such smoothed  fluctuations are independent of the number of dislocations $N$ and hence independent of $R$: $<Q^2>_ {smoothed}\rightarrow a +O (N^{-1})$. These two methods treat the medium beyond distance $R$ differently.\\

In previous work, we have shown that 2D and 3D infinite media \cite{Bart} reveal a diffuse behaviour as found by Wilkinson and Freund \cite{freund}. The role of the mean free path $\ell$ was also studied. In 3D, $<Q^2(R)>$ at large $R$ depends very weakly on the mean free path $\ell$, with a finite value $<Q^2(R)>/R$ for $\ell=\infty$. For 2D random media $<Q^2(R)>$ was seen to depend logarithmically on $k\ell$. In this paper we now consider a cylindrical wave guide which is a configuration used in several microwaves experiments by Sebbah \textit{et al.} \cite{Zhang,Genack2}. One additional reason for us to consider a wave guide is to have a genuine boundary. As a result the field is confined inside the wave guide so that we are sure not to forget any contribution for the screening process. Finally we will be able to study the influence of boundaries on the screening of topological charge. We study the dependence of $<Q^2>$ with radius $R$ and with mean free path $\ell$.\\

\section{Direct calculations}

We consider a complex random scalar wave field described by circular Gaussian statistics (known to be a good approximation for multiply scattered waves) \cite{Goodman,WRM}. The medium is a hollow conductive cylindrical wave guide of radius $R$ and with infinite length, containing disorder. The problem is formulated in cylindrical coordinates $(\rho,\theta,z)$. We impose that the field derivative cancels at the boundaries. In the following, we first present our calculation method of the topological charge variance $<Q^2>$.\\

\begin{figure}[!h]
\begin{center}
\subfigure[$\Psi_{2,0}(x,y)$]{\includegraphics[width=5cm]{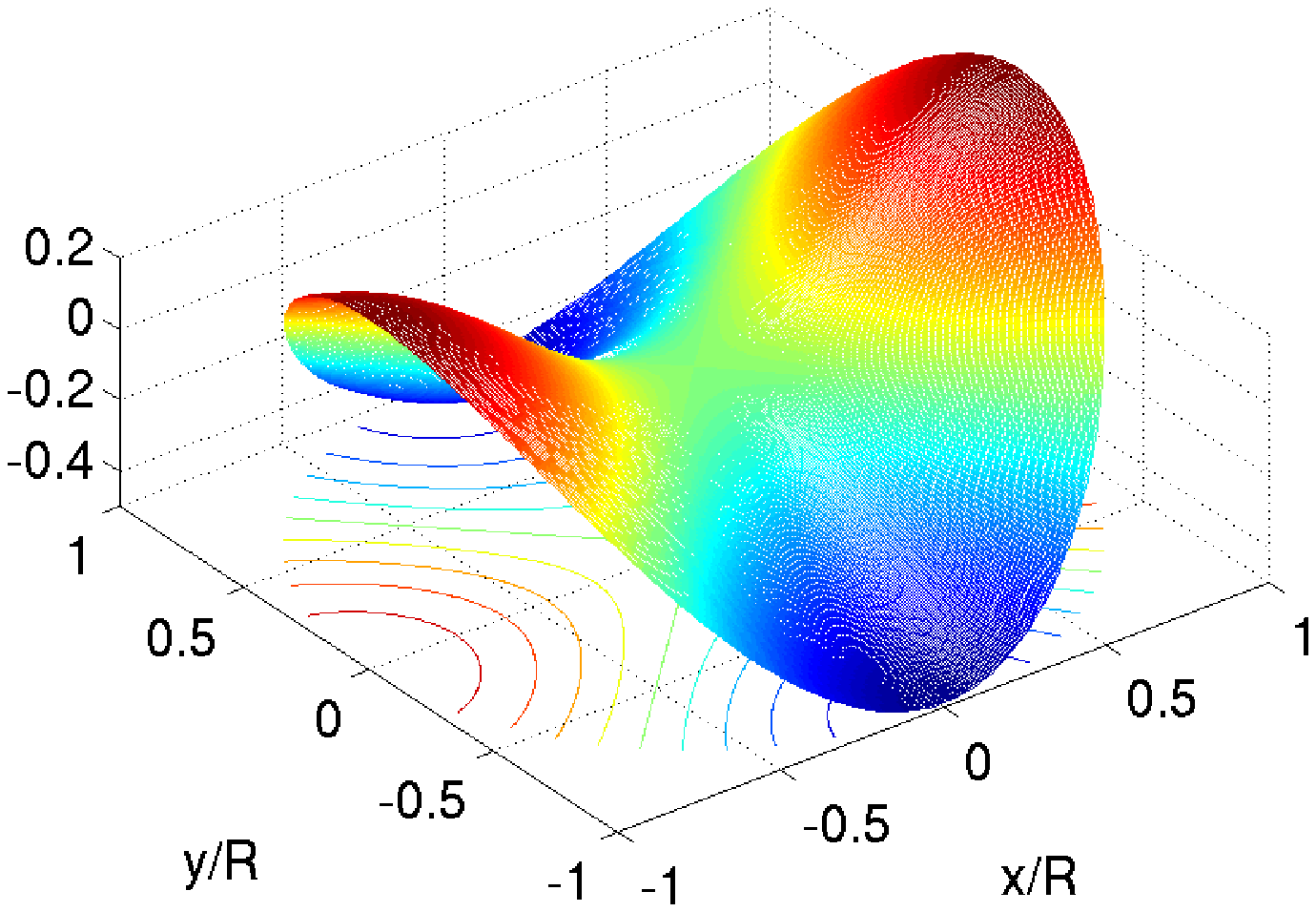}}
\subfigure[$\Psi_{6,0}(x,y)$]{\includegraphics[width=5cm]{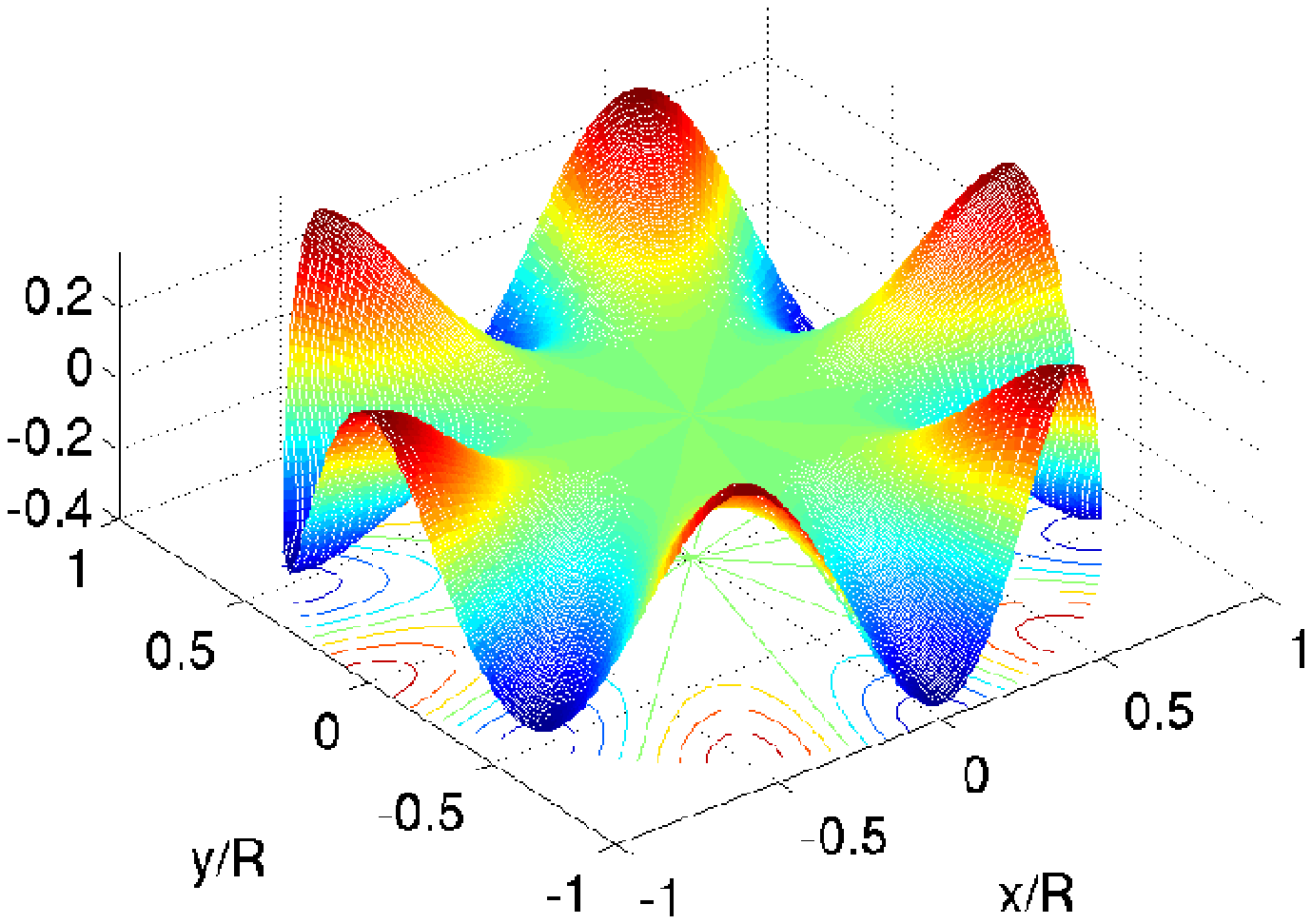}}
\subfigure[$\Psi_{19,0}(x,y)$]{\includegraphics[width=5cm]{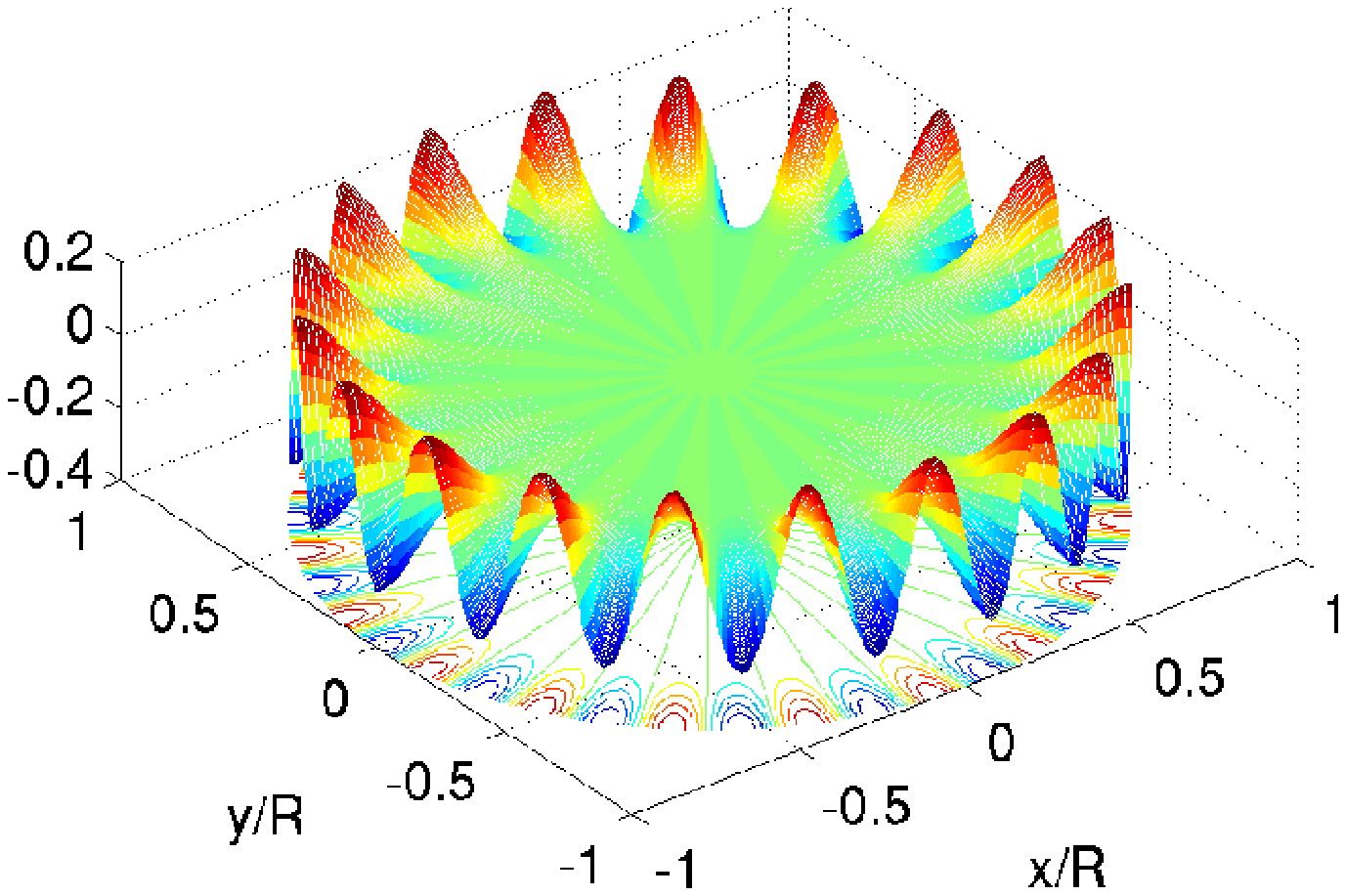}}
\subfigure[$\Psi_{6,2}(x,y)$]{\includegraphics[width=5cm]{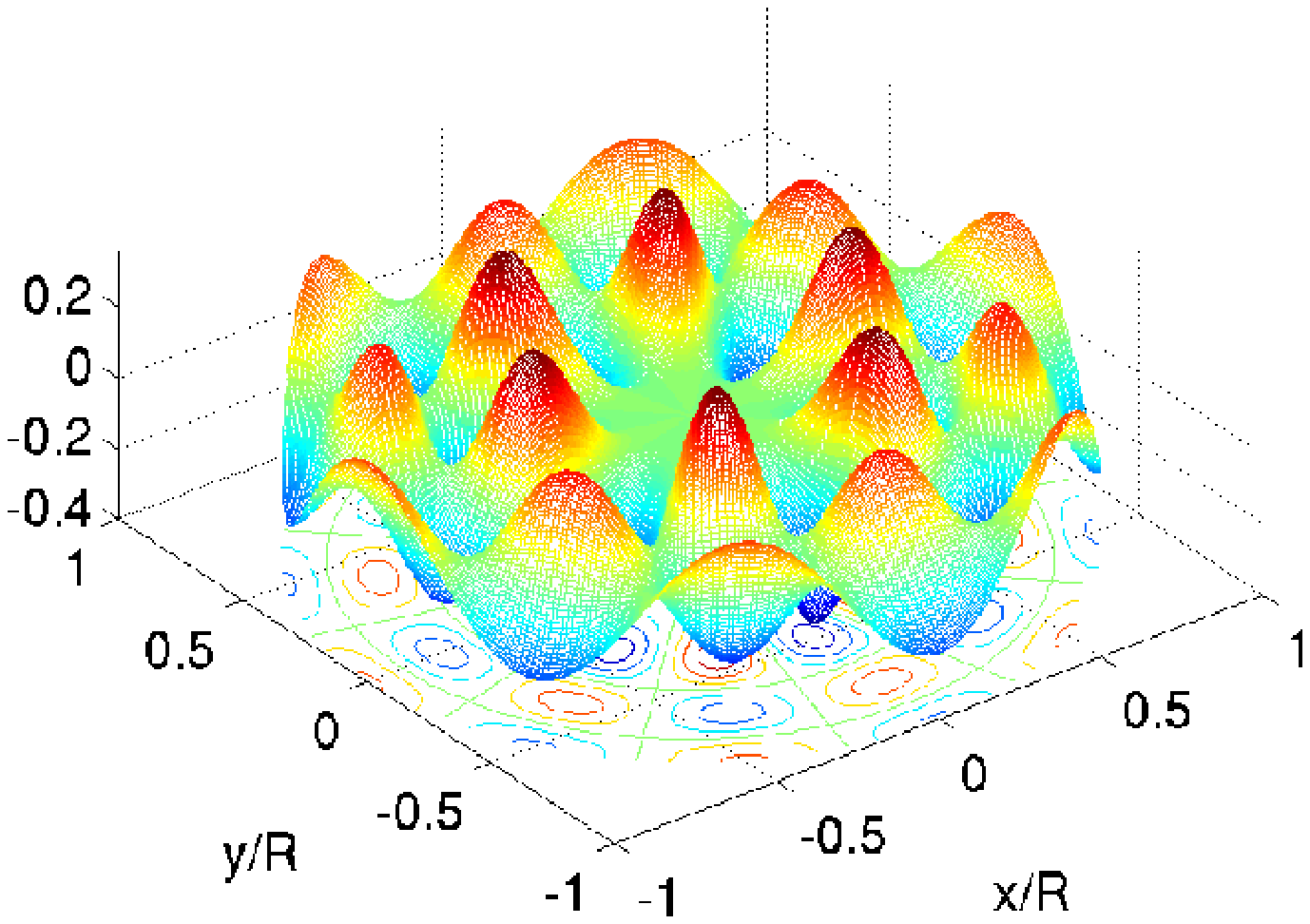}}
\caption{Examples of eigen modes $\Psi_{mn}$ in a cylindrical wave guide. The quantum number $m$ caracterises the dependence on the angular coordinate, and the quantum number $n$ caracterises the dependance on the radial coordinate: $\Psi_{mn}$ has $m$ radial node lines and $n$ circular node lines.}\label{Modes}
\end{center}
\end{figure}

The modes $\Psi_{mnk}$ (Fig. \ref{Modes}) of a homogeneous empty conducting cylindrical wave guide of radius $R$ and infinite length are given by:

\begin{equation}
\Psi_{mnk}(\rho,\theta,z) \propto J_m\left(\frac{\alpha_{mn} \rho}{R}\right) \exp(\mathrm i m \theta) \exp(\mathrm i kz)\,\,\,\,\mathrm{with}\,\,\,\, J_m'(\alpha_{mn})=0
\end{equation}
\noindent where $k\in\mathbb{R}$ is the wave vector, $m\in \mathbb{Z}$, $n\in \mathbb{N}$ and $\alpha_{mn}$ is the $n^{th}$ root of the first derivative  of the $m^{th}$ Bessel function. The dispersion relation for the mode $\Psi_{mnk}$ reads $\omega^2=k^2c^2+\alpha_{mn}^2c^2/R^2$ and has a cut-off frequency $\omega_{mn}=\alpha_{mn}c/R$. The modes $\Psi_{mnk}$ have a mean free path $\tau_{mn}(k)$.\\

\noindent In the presence of scattering, the averaged Green function can be written as \cite{Economou}: 
\begin{equation}
G^\pm(\rho,\theta,z;\rho',\theta',z';\omega)= \sum_{m,n}\int dk \frac{\Psi_{mnk}(\rho,\theta,z)\Psi^*_{mnk}(\rho',\theta',z')}{\omega_{mn}(k)^2-\omega^2\pm \mathrm i\epsilon+\mathrm i\omega/\tau_{mn}}
\end{equation}
where $\omega_{mn}(k)$ is the dispersion relation of $\Psi_{mnk}$ and $\tau_{mnk}=\ell_{mnk}/c$ is the life time of the mode $\Psi_{mnk}$. For simplicity we shall assume that all modes have the same life time $\tau =\ell/c$ although one can easily consider a more realistic model such that $\ell$ is different for all modes. The field correlation function $C_{\Psi}$ is proportional to the imaginary part of the Green function \cite{Green1}. Hence we can calculate the correlation function between two points separated in angle by $\Delta\theta=\theta'-\theta$ and located at same $\rho$ and $z$ from:

\begin{equation}\label{CorrField1}
\begin{array}{rl}
\displaystyle{C_{\Psi}(\rho,\Delta\theta,\omega)} & =<\Psi(\rho,\theta,z)\Psi(\rho,\theta',z)> \propto \mathrm{Im}\, G(\rho,\theta,z;\rho,\theta',z;\omega)\\
& \propto \displaystyle{\sum_{m,n} \frac{1}{R} \cos(m\Delta\theta) \frac{2 \alpha_{mn}^2}{\alpha_{mn}^2-m^2} \left| \frac{J_m(\frac{\alpha_{mn} \rho}{R})}{J_m(\alpha_{mn})} \right|^2 A_{mn}(\omega)}\\
\end{array}
\end{equation}

\noindent where the coefficients $A_{mn}(\omega)$ are given by:

\begin{equation}\label{CorrField2}
\displaystyle{\begin{array}{l}
A_{mn}(\omega)=\\
\left\lbrace \begin{array}{ll}
\sqrt{\frac{(\omega R/c)^2-\alpha_{mn}^2+\sqrt{((\omega R/c)^2-\alpha_{mn}^2)^2+(\omega R^2/\tau_{mn})^2}}{((\omega R/c)^2-\alpha_{mn}^2)^2+(\omega R^2/\tau_{mn})^2}}  & \mathrm{if}\,\,\,\frac{\omega R}{c} > \alpha_{mn}\\
\frac{\omega R^2/\tau_{mn}}{\sqrt{(\alpha_{mn}^2-(\omega R/c)^2)^2+(\omega R^2/\tau_{mn})^2}\sqrt{\alpha_{mn}^2-(\omega R/c)^2+\sqrt{(\alpha_{mn}^2-(\omega R/c)^2)^2+(\omega R^2/\tau_{mn})^2}}} & \mathrm{if}\,\,\, \frac{\omega R}{c} < \alpha_{mn}\\
\end{array}\right.
\end{array}}
\end{equation}

\noindent Note that for a finite mean free path $\ell$, the frequencies below the cut-off $\omega_{mn}$ contribute as well. For sufficiently large mean free path (in Fig.\ref{MFP} we will discuss how large $\ell$ must be), Eq. \ref{CorrField2} simplifies to:

\begin{equation}\label{CorrField}
\displaystyle{\begin{array}{c}
A_{mn}(\omega)=\left\lbrace \begin{array}{ll}
\displaystyle{\frac{1}{\sqrt{(\frac{\omega R}{c})^2-\alpha_{mn}^2}}} & \mathrm{if}\,\,\,\frac{\omega R}{c} > \alpha_{mn}\\
0 & \mathrm{if}\,\,\, \frac{\omega R}{c} < \alpha_{mn}\\
\end{array}\right.
\end{array}}
\end{equation}

\noindent For circular Gaussian statistics, the phase derivative correlation function $C_{\Phi'}(\Delta\theta)$ can be calculated from the field correlation function using the relation \cite{Bart}:
\begin{equation}\label{Corrprime}
C_{\Phi'}(\rho,\Delta\theta) = \left<\frac{d\Phi}{d\theta}(\rho,\theta)\frac{d\Phi}{d\theta'}(\rho,\theta')\right> = \frac{1}{2}(\log C_{\Psi}(\rho,\Delta\theta))''\log(1-C_{\Psi}^2(\rho,\Delta\theta))
\end{equation}

\noindent The variance $<Q^2(\rho)>$ of the total topological charge enclosed by a surface $s(\rho)=\pi\rho^2$ ($\rho\in[0:R]$) with a circular contour $\mathcal{C}(\rho)$ centered inside the wave guide is calculated from Stokes' theorem (Eq. \ref{Stokes}): 
\begin{equation}\label{var}
\begin{array}{rl}
\displaystyle{\langle Q^2(\rho)\rangle} & \displaystyle{=\frac{1}{(2\pi)^2}\oint_\mathcal{C}(\rho)\oint_{\mathcal{C}(\rho)}d\mathbf{s}\cdot \langle\boldsymbol{\nabla}\Phi(\mathbf{s})\boldsymbol{\nabla}\Phi(\mathbf{s}')\rangle\cdot d\mathbf{s'}}\\
& \displaystyle{=\frac{1}{(2\pi)^2}\oint_\mathcal{C}(\rho)\oint_{\mathcal{C}(\rho)}\rho^2d\theta d\theta'\left\langle\frac{1}{\rho^2}\frac{\partial\Phi(\rho,\theta)}{\partial\theta}\frac{\partial\Phi(\rho,\theta')}{\partial\theta'}\right\rangle=\frac{1}{2\pi}\int_0^{2\pi}d\Delta\theta\,C_{\Phi'}(\rho,\Delta\theta).}
\end{array}
\end{equation}

\noindent Eqs. \ref{var}, \ref{Corrprime} and \ref{CorrField1}-\ref{CorrField2}-\ref{CorrField} are calculated numerically. \\
 
 \setlength{\unitlength}{1cm}
\begin{figure}[!h]
\begin{center}
\subfigure[]{\includegraphics[height=5cm]{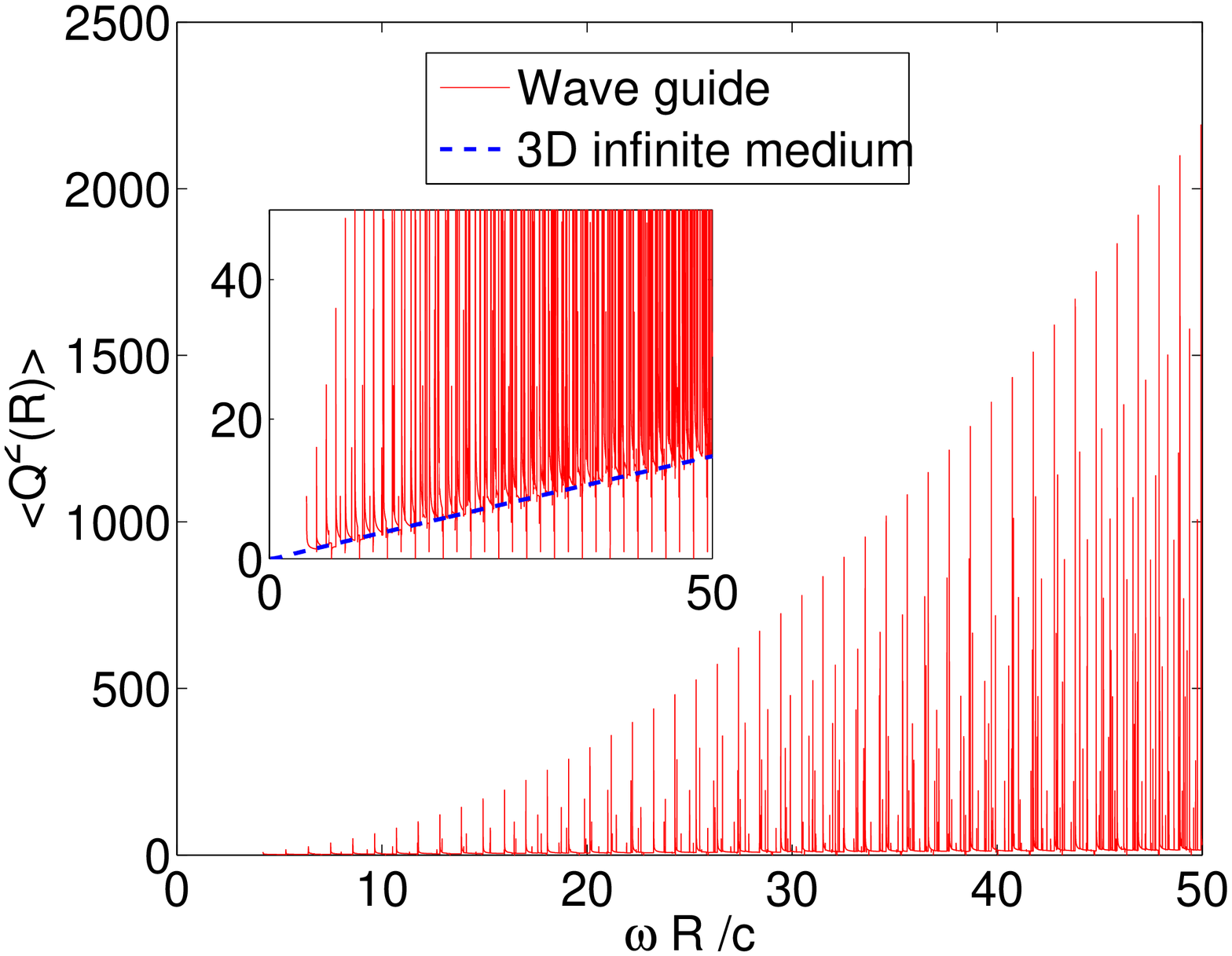}}
\subfigure[]{\includegraphics[height=5cm]{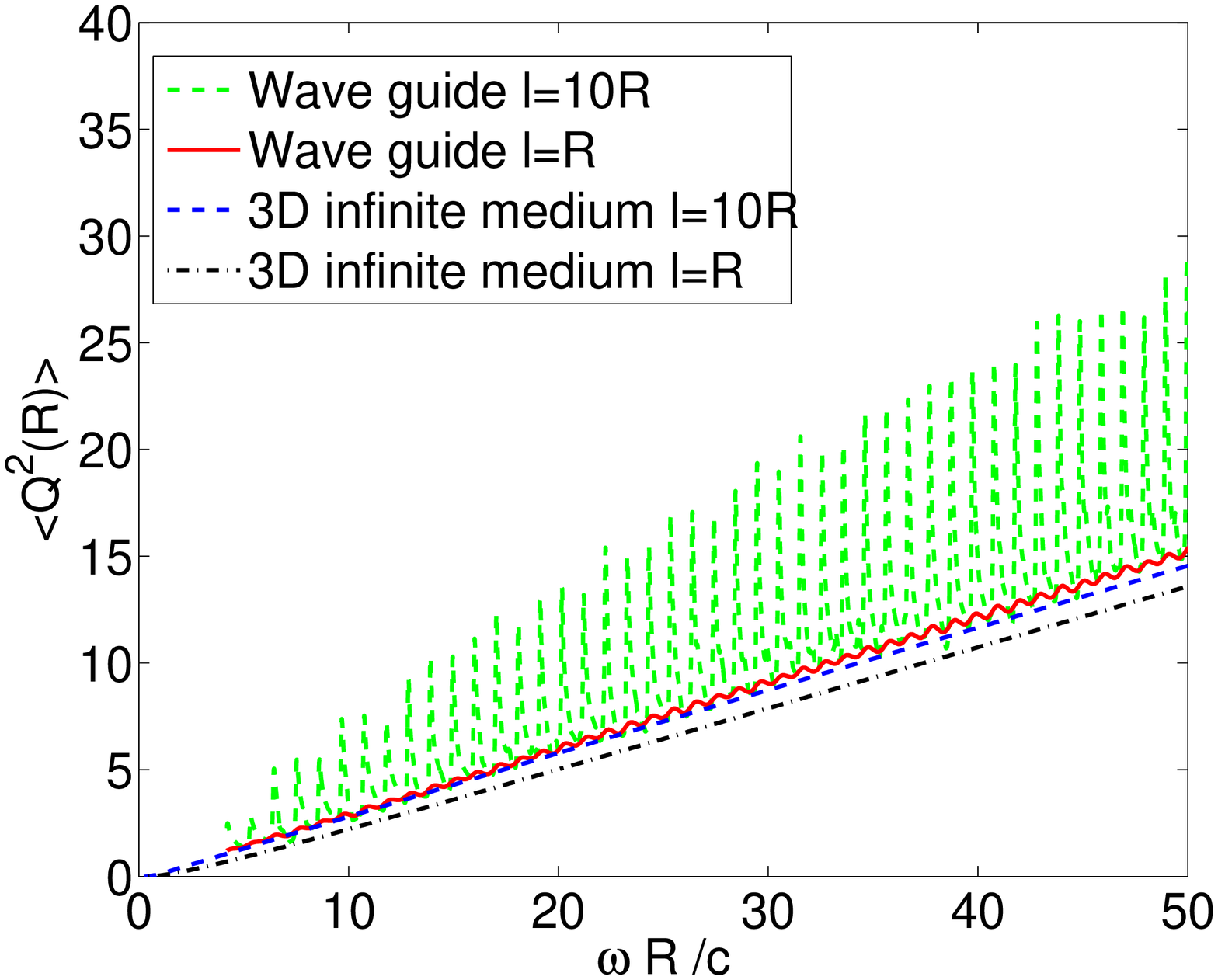}}
\caption{(color online) Evolution of the topological charge variance $<Q^2(R)>$ enclosed by a surface $\mathcal{S}=\pi R^2$ against the product $\omega R/c$, in red for a entire cross-section of the cylindrical conducting wave guide, in green dashed line for a 3D infinite medium. At the cut-off frequencies of the wave guide, sharp resonances arise. (a) In the limit of large $\ell$, calculated from Eq. \ref{CorrField}. Figure (b) shows the damping of the peaks with smaller $\ell$. (b) and the zoomed inset in (a) show that the global behaviour is close to the one predicted for a $3D$ infinite medium.}\label{Rezo}
\end{center}
\end{figure}
 
The results presented in Fig. \ref{Rezo} show that $<Q^2(\rho=R)>$ for the entire cross-section of the wave guide (red curve) tends to rise linearly with $\omega R/c$. This global behaviour is consistent with the calculation for a 3D infinite medium (green curve), with a diffuse behaviour revealing charge screening. On the other hand, near the cut-off frequencies, $<Q^2(R)>$ exhibits sharp resonances reaching a maximal value of $<Q^2(R)>=m^2$ for $\ell$ sufficiently large (i.e. when the limit of Eq. \ref{CorrField} is valid, figure 2a). For large $m$ this makes $<Q^2>$ much bigger than the prediction for a $3D$ infinite medium. We can explain the asymptotic peak value $m^2$ of $<Q^2(R)>$ with a simple argument. As we can see in Eq. \ref{CorrField}, when $\omega R/c\simeq\alpha_{mn}$ the averaged Green function is dominated by the two eigen modes $\Psi_{\pm m\, n}$ so that the field correlation function becomes $C(\theta,\rho)\simeq \cos(m\theta)$ which exhibits long range order. Then a short analytical calculation using equations \ref{Corrprime} and \ref{var} shows that $<Q^2>\simeq m^2$. Note that we study the limit $\ell\rightarrow\infty$ only with a pedagogical purpose; a truly homogneous wave guide would not exhibit Gaussian statistics.\\

For a finite mean free path $\ell$ (Fig. 2b) the resonant peaks of $<Q^2>$ are strongly attenuated though still visible until $\ell\geq R$. The disorder introduces a coupling between the modes and the relative importance of the eigen modes diminishes so that the other modes start to contibute to the field correlation function $C_{\Psi}$. Fig. \ref{MFP} shows the slow dependence of the peak value of the topological resonance on the mean free path $\ell$. Upon varying  10 orders of magnitude over the mean free path the fluctuations vary only by 2 orders of magnitude. The maximum value $<Q^2(R)>=m^2$ is reached only for very large mean free path and for $\ell\simeq 100R$ the peak value is already suppressed by a factor of $10$.\\

\begin{figure}[!h]
\begin{center}
\includegraphics[width=8cm]{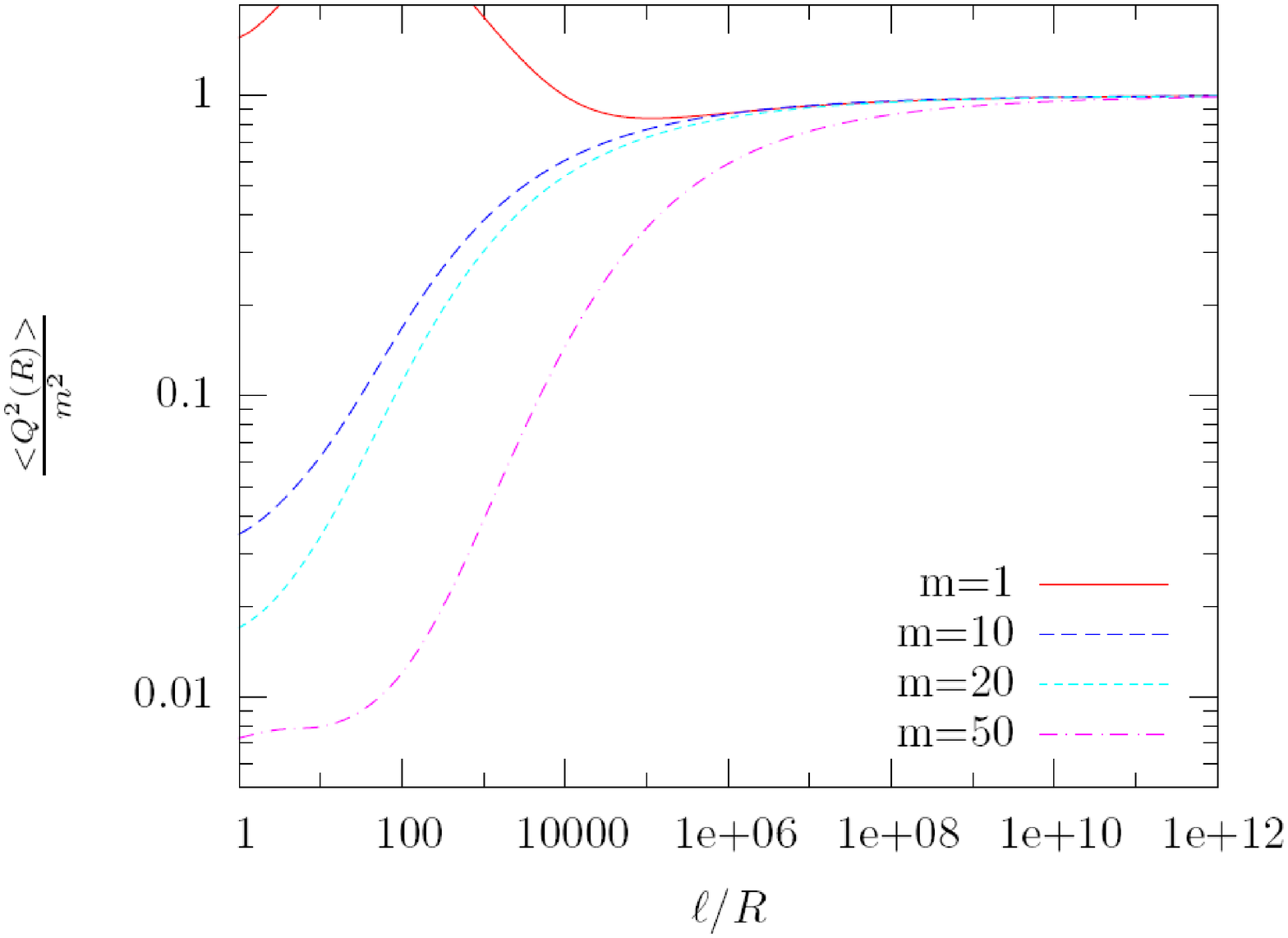}
\caption{(color online) Semilogarithmic plot of $<Q^2(R)>/m^2$ at different dominant cut-off frequencies (for $n=0$ and for different $m$) as a function of the mean free path normalised by the wave guide radius. For $\ell \lesssim R$, $<Q^2>$ approaches the value predicted for a $3D$ infinite mediu. Only for $\ell$ as large as $10^9$, the maximum value $<Q^2>\sim m^2$ is reached.}\label{MFP}
\end{center}
\end{figure}

\begin{figure}[!h]
\begin{center}
\subfigure[$n=0$]{\includegraphics[width=6.5cm]{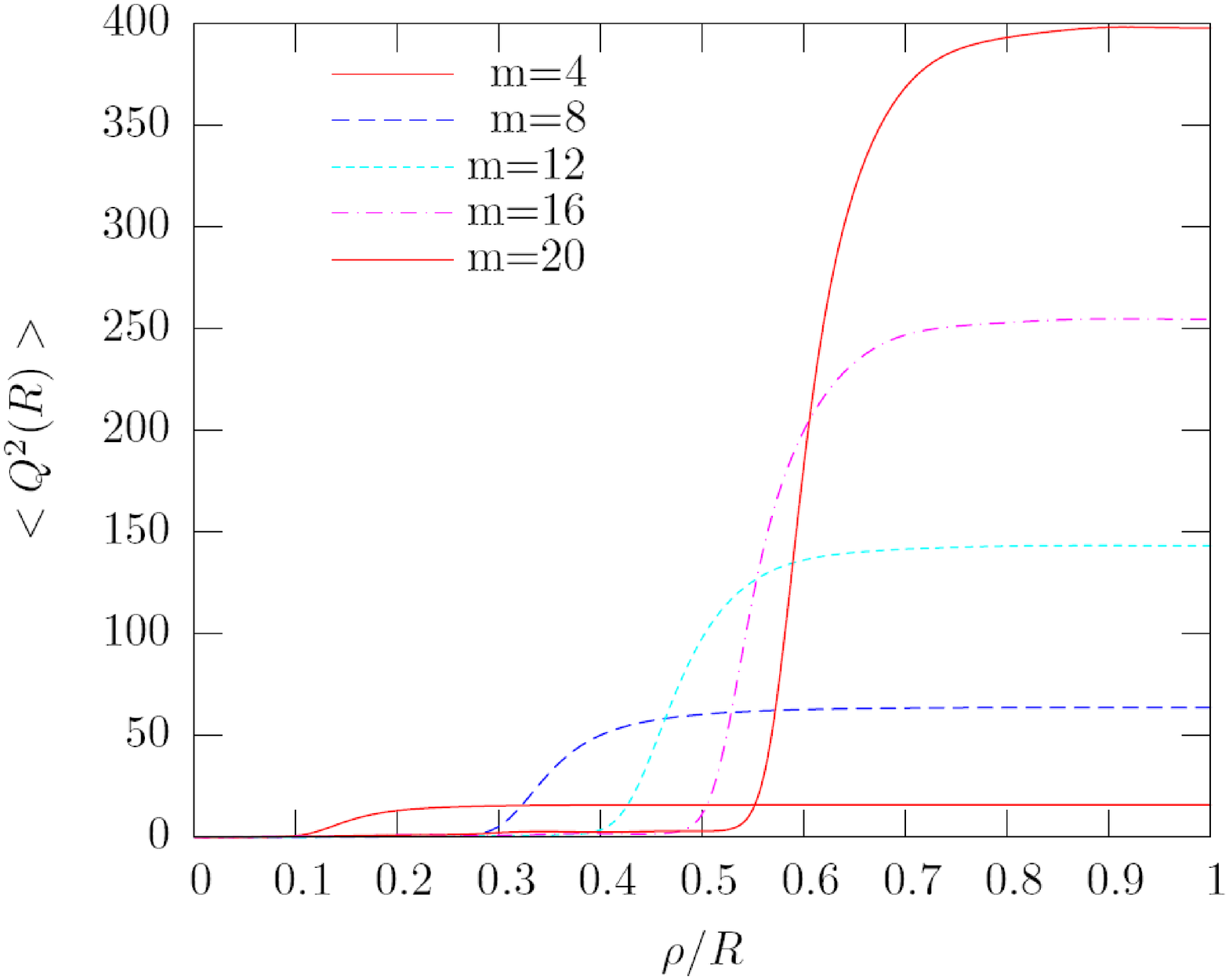}}
\subfigure[$m=19$]{\includegraphics[width=6.5cm]{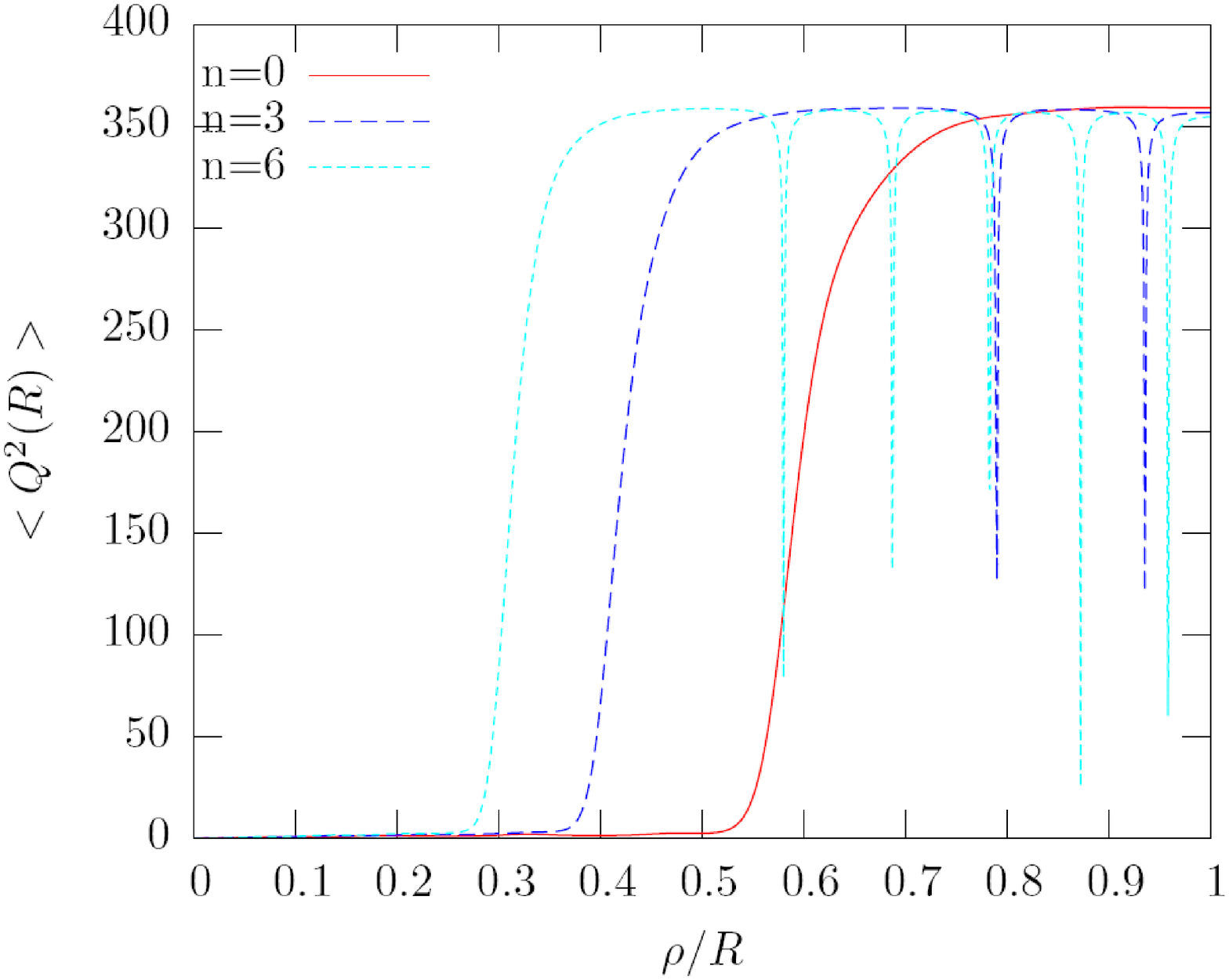}}
\caption{(color online) Evolution of the variance of the topological charge $<Q^2_{mn}(\rho)>$ enclosed by a disk of radius $\rho$ centered inside the wave guide at different cut-off frequencies $\omega_{mn}$ and for $\ell$ sufficiently large (a) at fixed $n=0$ and for different $m$ (b) at fixed $m=19$ and for different $n$. We can see that there is a critical radius $\rho_c$. For $\rho<\rho_c$, charges are screened and for $\rho>\rho_c$ charges of the same sign accumulate to give huge $<Q^2_{mn}(\rho)>$. $\rho_c$ tends to increase with $m$ and decrease with $n$.}\label{Radius}
\end{center}
\end{figure}

In Fig. \ref{Radius}, we study how $<Q^2_{mn}(\rho)>$ at the cut-off frequency $\omega_{mn}$ varies with the surface $s=\pi\rho^2$ (with $\rho\in[0:R]$) centered inside the wave guide. This calculation reveals that the large value for $<Q^2(R)>$ is essentially an effect localized near the edge.  The charges are screened in the center of the wave guide but not near the edges. Note that when $m$ increases and $n$ decreases , the unscreened charges localise closer to the edges.\\

\section{Simulation}

The direct calculation of topological charge fluctuations reveals a linear dependence of the charge variance on the surface raduis caused by charge screening, and special frequencies for which screening seems absent. For the dominant peaks ($n=0$), $\alpha_{m0}\sim m$ and hence $<Q^2(R)>=m^2\sim \alpha_{mn}^2\sim (\omega_m R/c)^2 \propto R^2$ scales with the surface as if all the charges were independent. It is also seen in Fig. \ref{Radius} that these fluctuations mainly come from the edges of the wave guide and that the peak value of the resonances depends on the number of radial node lines $m$ and not on the number of angular node lines $n$. We get deeper insight into this phenomenon using a computer simulation that generates a random circular Gaussian complex wave field in a transverse cross-section of the wave guide. To this end we write: $\Psi(\rho,\theta)=\sum_{mn}a_{mn}\Psi_{mn}(\rho,\theta)$. Here $\Psi_{mn}(\rho,\theta)$ are the normalised eigen modes and $a_{mn}$ are randomly generated coefficients obeying a Gaussian circular statistics with a variance fixed by:
$<a_{mn}^2>=A_{mn}(\omega)$ (to be consistent with Eq. \ref{CorrField1}).\\

\begin{figure}[!h]
\begin{center}
\subfigure[]{\includegraphics[height=6cm]{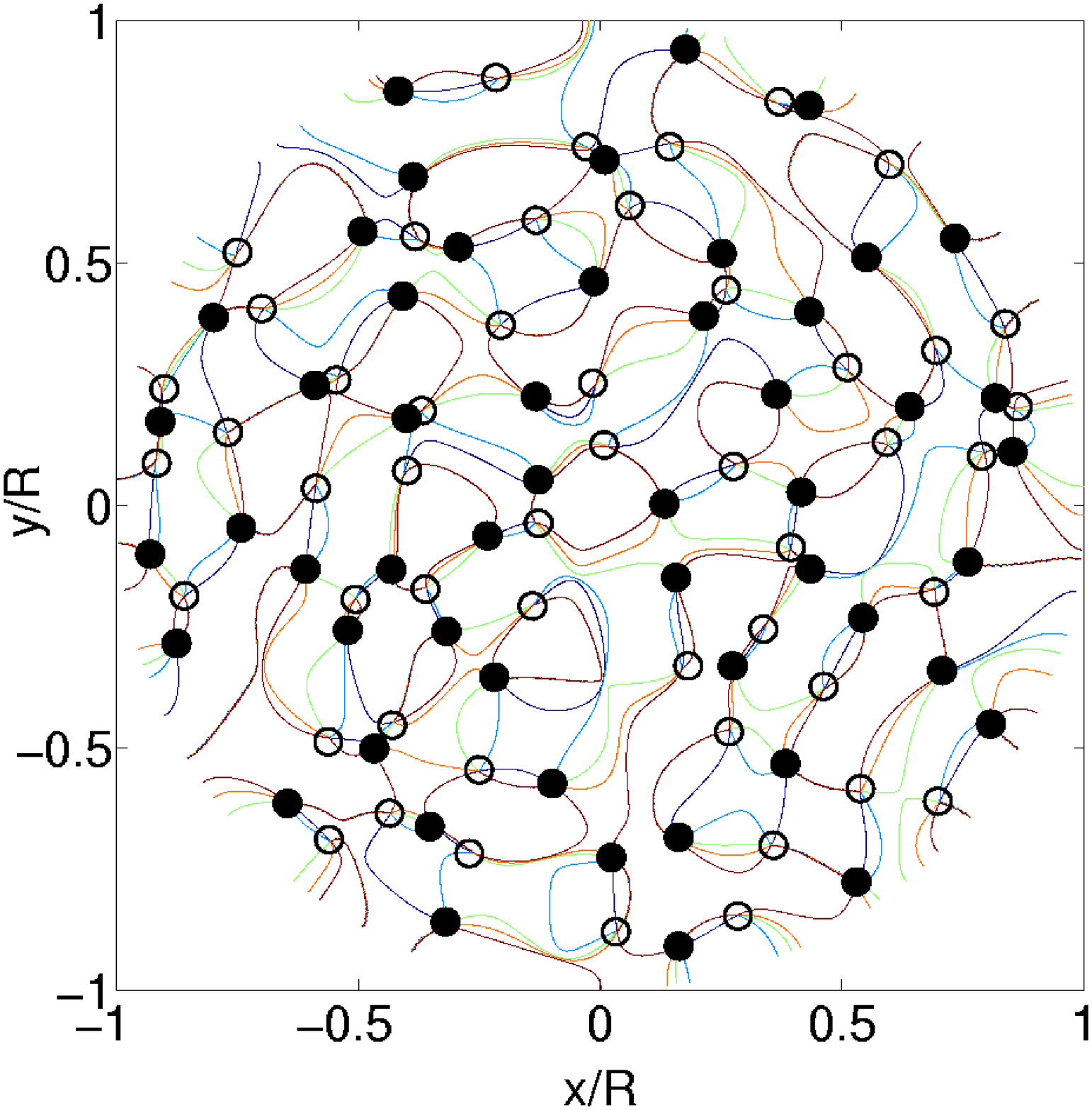}}
\subfigure[]{\includegraphics[height=6cm]{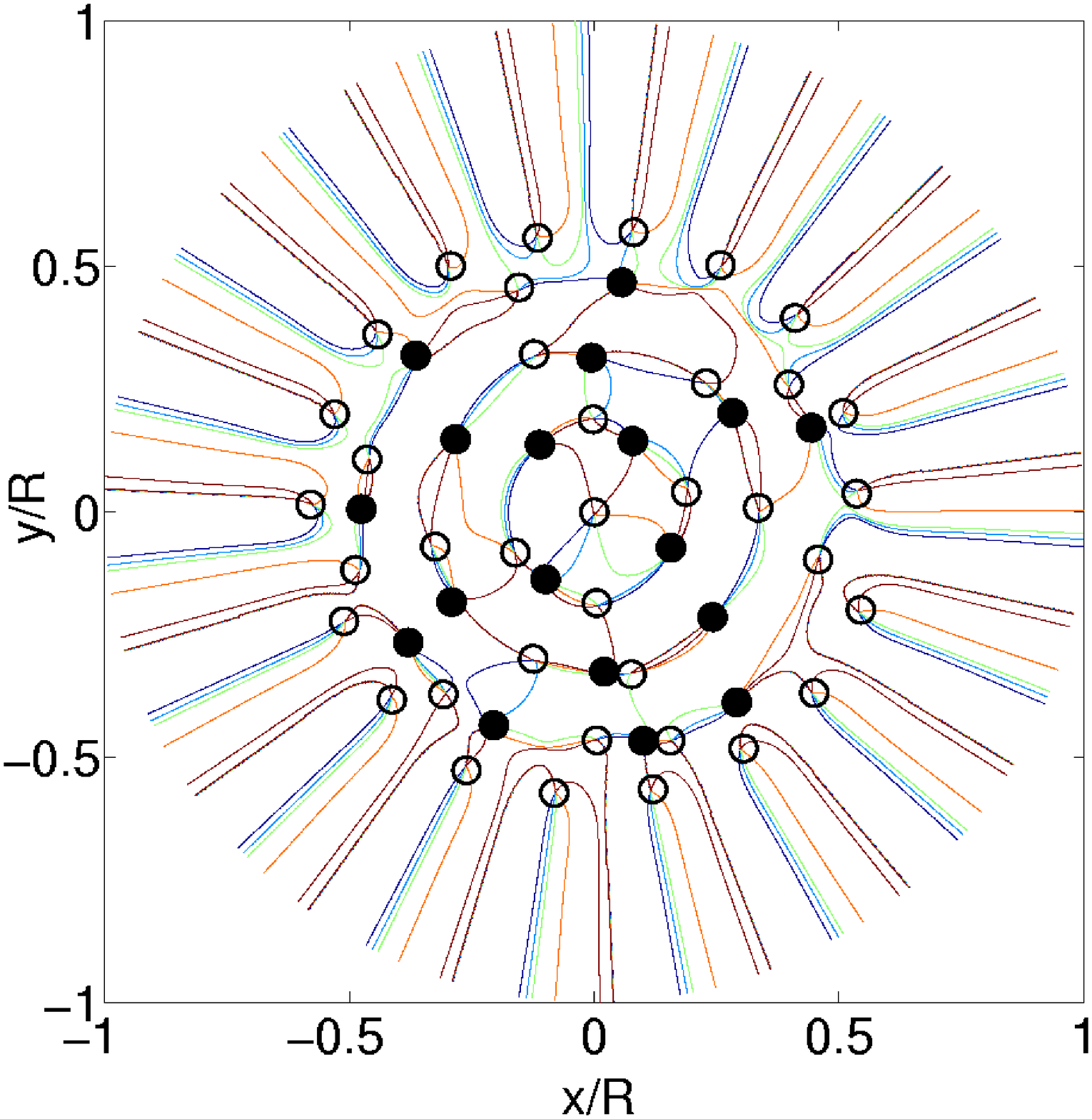}}
\subfigure[]{\includegraphics[height=6cm]{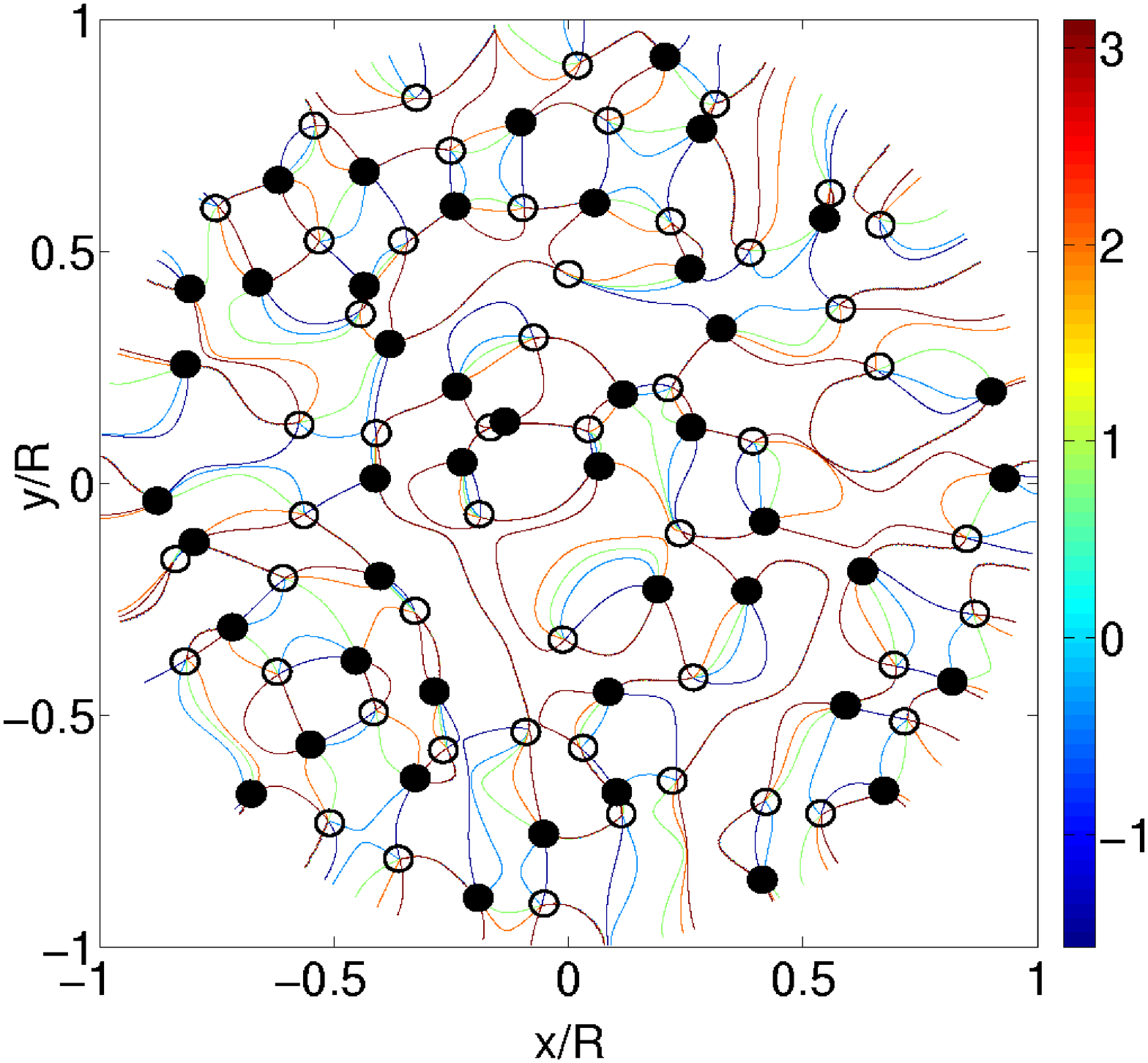}}
\caption{(color online) Representation of one realisation of the random field in the wave guide, for different $\ell$ and for different $ \omega R/c$. ($\bullet$) denotes a phase singularitiy with negative charge, ($\circ$) denotes a phase singularity with positive charge. The colored lines denote equiphases. (a) Non cut-off frequency $\omega R/c=20.72$: $Q=-2$; (b) Cut-off frequency $\omega R/c \gtrsim \alpha_{19,0}=21.18227$ with $\ell\rightarrow\infty$: $Q=19$; and (c) Same frequency $\omega R/c \gtrsim \alpha_{19,0}=21.18227$ but now with a mean free path $\ell=10R$: $Q=5$}\label{Equiphases}
\end{center}
\end{figure}

Fig. 5a shows the expected screening away from the cut-off frequencies. Fig. 5b shows that close to a cut-off frequency, charges tend to be screened in the center but exhibit the same sign near the edges consistently with the result obtained in Fig.\ref{Radius}.\\

The twin principle \cite{twin} imposes that if the field is continuous each topological charge is necessarily connected by equiphase lines to a singularity of opposite sign. This generates the usual charge screening observed in the center, similar to infinite media. However in the presence of boundaries the twin principle no longer holds and isolated singularities may be created or destroyed at a boundary. This perturbs the balance between positive and negative charges. Consequently, a huge topological charge can appear in a circular area without an important change in the number of singularities. Away from the cut-off frequencies, these isolated singularities are independent of each other and their number is proportional to the perimeter so that we find $<Q^2(R)>=N_{\textit{Boundary}}\propto R$. However at the cut-off frequencies $\omega=\alpha_{mn}c/R$ the field is dominated by the weight of the two eigen modes $\Psi_{\pm m,n}$ with $m$ radial nodal lines that exhibit $m$ isolated singularities. The larger $m$, the more equiphase lines end up at the boundaries and the more singularities have become isolated. This increases the probability to have many singularities of the same sign, and this probability thus increases the variance $<Q^2(r)>$. As we can see in Fig. 5b these boundary singularities are not independent (which would give $<Q^2>\rightarrow m$) but tend to be of the same sign so that $<Q^2>\rightarrow m^2$ for a large mean free path. The huge fluctuations at special frequencies are thus not due to independent charges in the total surface $\mathcal{S}$ but to collective effects near the edges.
 
\section{Conclusion}

A superposition of waves scattered by a disordered medium gives rise to a speckle pattern which presents a complicated network of phase vortices. We have studied the role of mean free path and boundaries on the screening between the topological charges of the phase vortices. The same linear diffuse law that relates fluctuations of topological charge and enclosed surface, as already found for $3D$ infinite media, is seen. However, at the cut-off frequencies of the wave guide, giant fluctuations of the topological charge occur. These fluctuations are very sensitive to the disorder and probe the mean free path even when it is much larger than the wave guide size.

\end{document}